# On the impact of the geospace environment on solar-lithosphere coupling and earthquake occurrence

**Dimitar Ouzounov**[1,*] and **Galina Khachikyan**[2]

[1] Center of Excellence in Earth Systems Science & Observations, Chapman University, Orange, CA, USA, ouzounov@chapman.edu
[2] Institute of Ionosphere, Almaty, Kazakhstan, galina.khachikyan@gmail.com
[*] Correspondence: ouzounov@chapman.edu;

**Abstract:** We have found that about two months after creating a new radiation belt in the inner magnetosphere due to a geomagnetic storm, an increasing seismic activity may occur near the magnetic field lines' footprint of a newly created radiation belt. The Combined Release and Radiation Effects Satellite (CRRES) detected a new radiation belt after a geomagnetic storm on March 24, 1991. Shortly after that, on May 30, 1991, a strong M7.0 earthquake occurred in Alaska in the footprint of geomagnetic line L~2.69. Additionally, on October 28, 2012, a strong M7.8 earthquake occurred in Canada near the footprint of L~3.3, which was close to the magnetic lines of a new radiation belt detected by a satellite "Van Allen Probes" after a geomagnetic storm on September 3, 2012.

Seismic activity also increased near the magnetic field lines' footprint of a newly created radiation belt around L~1.5-1.8 due to a geomagnetic storm on June 21, 2015. We demonstrate the possible existence of two way of solar-lithosphere coupling processes : (i) the disturbances in the lithosphere, accompanying the earthquake preparation process, can modify the electric field in the global electric circuit (GEC), which results in appearing of disturbances in the ionosphere; and the vice-versa mechanism (ii)  the solar wind-generated disturbances in the magnetosphere and ionosphere, can modify the electric field in the GEC, that will result in appearing of disturbances in the lithosphere
.

**Keywords:** earthquakes, cosmic rays, geomagnetic storms, radiation belt, LAIC

## 1. Introduction

An idea that space weather, as measured by sunspots, cosmic rays, solar wind, interplanetary magnetic field, geomagnetic activity, and precipitation of charged particles from the radiation belt, may play a role in triggering earthquakes, has a long story but did not receive a physical justification for the time being. It was noted in [1] that the difficulty of solving this problem is that the unknown is a physical mechanism of action of relatively weak fields of cosmic origin on compelling tectonic processes. So, the authors of [2] have considered the thermal mechanism, in which the current Foucault, induced in the Earth's crust by a variable magnetic field, led to the additional heating of the mountain breed, but as marked in [1], it is not clear how such an insignificant heat affects the probability of the occurrence of earthquakes. It was also assumed that the action could be forceful, caused by the movement of telluric currents in the Earth's magnetic field. However, in work [3], it was shown that arising from these ponderomotive forces causes only the weakest stresses and deformations in the Earth's crust. As suggested authors of [4, 5], the Alfvèn waves, which can drive currents in the ionosphere, which then reradiate the energy as electromagnetic waves that propagate to the ground [6], could mediate space weather with seismic activity. Here one may remember that over the last 30 years, the international community's cooperation has already suggested a physical mechanism for the ionosphere-lithosphere coupling, while for the opposite process: for the propagation of the electromagnetic disturbances generated in the lithosphere during an earthquake preparation, to the ionosphere. This is a concept of the lithosphere-atmosphere-ionosphere coupling (LAIC), Here one may remember that over the last 30 years, the international community's cooperation has already suggested a physical mechanism for the ionosphere-lithosphere coupling, while for the opposite process: for the propagation of the electromagnetic disturbances generated in the lithosphere during an earthquake preparation, to the ionosphere. This is a concept of the lithosphere-atmosphere-ionosphere coupling (LAIC), which shows that disturbances in the ionosphere are caused by modifying the electric field in the global electric circuit (GEC) by electromagnetic disturbances in the lithosphere accompanying the earthquake preparation process [7-11, and references in herein]. The concept of the GEC was successfully used in [12, 13] to explain the results from the DEMETER satellite observations [14, 15], which showed a statistically significant decrease in the natural VLF (~1.7 kHz) wave intensity related to 8400 nighttime earthquakes with M ≥ 5.0 within 440 km of the epicenters.

The classical concept of GEC, initially proposed in [16], is a system of stationary currents between conducting Earth and the ionosphere. At present, a modified configuration of the GEC is discussed [17,18]. Its external element is located on the magnetopause with the electromotive force generator driven by solar wind energy in a modified GEC. Its internal element is located in the solid Earth with the electromotive force generator driven by the tectonic processes [18]. Considering this, one may suggest that the modified GEC may be considered a mediator in the transmission of solar wind energy into the Earth's crust.

The functioning of a GEC will depend on electrical conductivity along the entire path from the magnetopause to the Earth's crust. Along this path, the mesosphere, stratosphere, and troposphere have problems with air ionization and, thus, with conductivity. The sources of air ionization in these layers may differ with different Spatio-temporal variations of their characteristics. This may lead to the appearance of different Spatio-temporal variations in seismicity characteristics.

So, in the near-ground troposphere, the ionization may be produced by the isotopes of Radon [9,11,19,20], which concentration shows relatively strong spatial variations and differs widely above continental and oceanic areas. An average radon flux density for the entire Ocean equal to 0.0382 mBq m$^{-2}$ s$^{-1}$ [21] is much smaller than the typical estimates for the average flux density from land, which are in the approximate range of 20 to 35 mBq m$^{-2}$ s$^{-1}$ [22]. Over the land, the radon concentration depends on the tectonic conditions. For example, in Mysore city (India), it is of order 20 Bq/m3 [23]; simultaneously, in the vicinity of active faults, radon concentration may reach 2000 kBq/m3, five orders of magnitude larger [24]. The above means that the ionization of the near-ground troposphere and electrical conductivity here may show relatively strong spatial variations that will influence the spatial functioning of GEC and seismic activity (in the frame of our suggestion). The strongest seismic activity is expected to be in the vicinity of active faults, which indeed takes place. In the upper troposphere and lower stratosphere, the galactic cosmic rays influence the ionization and, therefore, the electrical parameters of the atmosphere [25, 26]. The ionization due to galactic cosmic rays (GCR) always exists in the atmosphere, and it changes with the 11-year solar cycle due to solar modulation. The intensity of the GCR increases in solar minimums, resulting in favorite conditions for the GEC's functioning and seismic activity increasing in solar minimums. In addition to continuous ionization in the Earth's atmosphere caused by a galactic cosmic ray, a sporadic ionization occurred during solar energetic particle events, potentially affecting the Earth's environment [27, 28] that also could input to GEC operation and earthquake occurrence. In subsection 3.1, we analyze the temporal variations in 1973 – 2017 the daily M≥4.5 earthquake counts and daily amount of released at the globe seismic energy by temporal variations of solar activity as measured by the daily mean sunspot numbers. Due to geomagnetic storms, the high-energy electrons filling the outer radiation belt can spill down and populate the underlying geomagnetic lines forming the new radiation belts (storage ring) around certain geomagnetic lines in the inner magnetosphere, which can exist from several days to several tens of months [29-38]. Since a storage ring is forming at specific geomagnetic lines, one may suggest the air conductivity could be increased precisely along these lines, and when they cross the regions of the Earth's crust and where the tectonic stresses are close to the threshold of the destruction of the rock, earthquakes might occur at the base of these lines. Such earthquakes may be considered as the targeted ones. We have attempted to verify this assumption and present obtained results in subsection 3.2. On the definition, the input of corpuscular space energy into the Earth's environment is controlled by the main geomagnetic field. Thus, if corpuscular space energy is input to seismic activation, one may suggest the existence of geomagnetic control of seismicity. In subsection 3.3, we present some results which show that this is a case.

## 2. Data and Methods

In this work, we used data of the USGS global seismological catalog (https://earthquake.usgs.gov/earthquake) for 1973-2017 for earthquakes with a magnitude of M≥4.5 (more than 220 thousand events). For epicenters of all analyzed earthquakes, we calculated parameters of the main geomagnetic field and values of the McIlvaine parameter – L [39], indicating the distance of geomagnetic lines to the center of the Earth at the equator expressed in the radii of the Earth, using the IGRF model and the computer codes of the GEOPACK program [40].

A geomagnetic storm is an interval of several days duration during which there is a significant reduction in the horizontal component of the geomagnetic field at the Earth's surface [41]. Following [42], geomagnetic storms may be small (Dst from −30 to −50 nT), moderate (Dst from −50 to −100 nT), strong (Dst from −100 to −200 nT), powerful (Dst from −200 to −350 nT), and extra strong (giant) with Dst – index lower of −350 nT. To investigate a response of seismic activity to geomagnetic storms, the data on the Disturbance Storm (Dst – index) were taken from NASA/GSFC's Space Physics Data Facility's CDAWeb service and OMNI data (https://cdaweb.gsfc.nasa.gov/index.html/). Geomagnetic storms, the bright manifestation of space weather variations, devastate the Earth's radiation belts (toroids of very high-energy magnetically trapped charged particles) discovered in 1958 by Van Allen [43]. Radiation belts typically comprise two distinct zones (inner and outer) spatially separated by the slot region (Figure 1 from [29]).

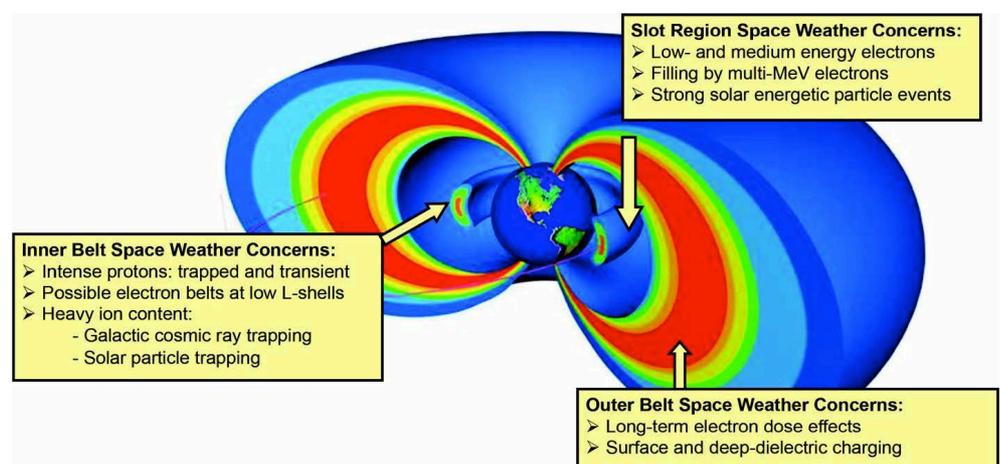

**Figure 1.** Schematic diagram of Earth's radiation belt from [29].

The inner Van Allen zone, mainly populated by the energetic protons, extends from just above the dense atmosphere out to an equatorial altitude of about 10,000 km above the Earth's surface. The leftmost label in Figure 1 shows that the space weather concerns for the inner radiation belt zone are several [29]: the intense, high-energy trapped protons, the variable, trapped solar energetic particles, the trapped galactic cosmic rays, and the trapped energetic electrons. The upper label in Figure 1 shows that the slot region can present several space weather concerns, including low- and medium-energy electron enhancements, multi-MeV electrons (on rare occasions), and strong solar energetic particle events (again on relatively rare occasions). The slot region extends from L ~ 2.0 to L ~ 3.0. The outer zone of the radiation belt extends from L ~ 3.0 to L ~ 6.5. It comprises mildly to highly relativistic electrons and varies widely in time and particle intensity [29]. Due to geomagnetic storms, the high-energy electrons can spill down from the outer radiation belt and form a new radiation belt (storage ring) around certain geomagnetic lines in the inner magnetosphere [29-38]. The new radiation belts exist mainly for several days [38], but very rare they may exist for up to several weeks and months. We analyze the response of seismic activity at three widely discussed geomagnetic storms, which were followed by a newly forming long-living radiation belt in the inner magnetosphere: 24 March 1991 [33], 1 September 2012 [35], and 21 June 2015 [36].

To correlate seismic data with solar activity variations, we used daily sunspot numbers from the World Data Center SILSO, Royal Observatory of Belgium, Brussels (**https://wwwbis.sidc.be/silso/datafiles**).

### 3. Results

3.1 *On temporal variations of solar activity and global seismicity*

In Figure 2, the lower panel shows variations in 1973-2017 the daily mean sunspot numbers for 21 - 24 solar cycles; the middle panel presents variations of a daily number of earthquakes with M≥4.5 occurring on the globe, and the upper panel shows the logarithm of the daily amount of released at the globe seismic energy (Log Es = 1.5 $M_W$ + 11.8) in Joules.

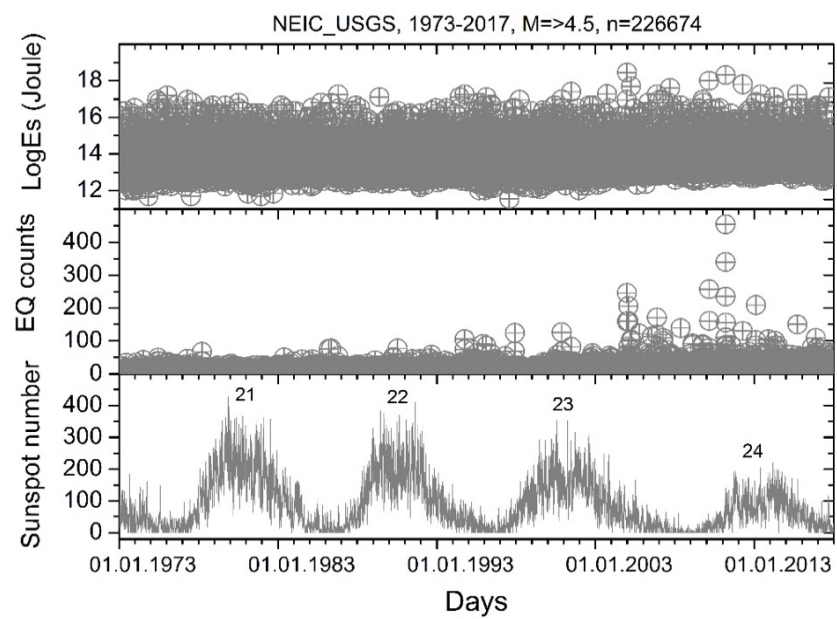

**Figure 2.** (lower panel) – Variations in 1973-2017 the daily mean sunspot numbers for 21 - 24 solar cycles; (middle) – variations of a daily number of earthquakes with M≥4.5 occurred on the globe, and (upper panel) - variations of the logarithm of the daily amount of released on the globe seismic energy Log E in Joules

Figure 2 shows that a noticeable increase of daily mean earthquake counts (middle panel) and released seismic energy (upper panel) took place after the start of the 21st century. During that time, the clustering of strong earthquakes occurred on the globe [46-50]. Namely, three strong events were in Indonesia near the island of Sumatra in 2004. M9.1, in 2005. M8.5, and in 2012, M8.5; three in Chile in 2010. M8.8; in 2014, M8.2, and in 2015. M8.3; two on the Kuril Islands in 2006, M8.3, and in 2007, M 8.1; in Japan in 2011, M9.0; the Sea of Okhotsk in 2013, M8.3; Mexico City in 2017, M8.2. This cluster of seismic events resulted in Figure 2 in increasing LogEs and earthquake counts during the declining phase of the 23 solar cycles and developing of 24 solar cycles.

It is seen from Figure 2 that the intensity of the 11year solar cycles varies over time. The state of solar activity, in which the intensity of several consecutive 11year cycles is significantly less than the average value, is characterized as a grand solar minimum. As predicted in [44], a new solar grand minimum began to develop from the beginning of the 21st century. The previous solar grand minimum, named after the astronomer Gleisberg, took place from 1880 to 1915 [45].

During the Gleisberg solar grand minimum, the temporal clustering of the strongest (M≥8) earthquakes also occurred. These events occurred on the Tien Shan in 1911, M8.2; Alaska in 1899, M 8.0; Western Turkmenistan in 1895, M 8.0; Kashgariya in 1902, M8.2; Northern Mongolia in 1905, M8.2; California in 1906, M8.3; China in 1906, M8.3; Colombia in 1906, M8.6 [49].

Solar cycles modulate the flux of Galactic Cosmic Rays (GCR) out of phase. Therefore, the grand solar minimum will "produce" the GCR's grand maximum, which could increase air ionization and electric conductivity in the troposphere-stratosphere. This, in turn, will increase the efficiency of GEC functioning – one of the possible mediators in the transmission of solar wind energy into the Earth's crust following a modified conception of the GEC [17, 18].

Figure 3a shows the distribution of the logarithm of the daily amount released on the globe seismic energy (LogEs) in dependence on daily mean sunspot numbers, where the red line is a linear trend. Figure 3b shows separately linear trend of Log Es versus sunspot numbers (SSN) as follows: $LogEs = 13.8 - 4.99\ SSN^{-4}$, with $R^2 = 0.003$, SD = 0.7.

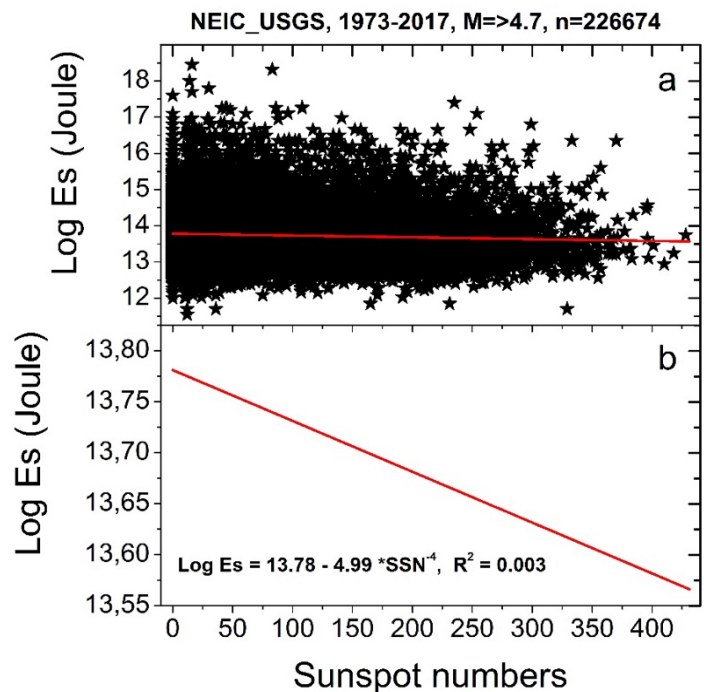

**Figure 3.** (a) – The distribution of the logarithm of the daily amount of released on the globe seismic energy (LogEs) in dependence on daily mean sunspot numbers, the red line is a linear trend; (b) –the linear trend of LogEs versus SSN as follows: Log Es = 13.8 – 4.99 SSN$^{-4}$, with R$^2$ = 0.003.

In 1973-2017, the daily LogEs values ranged between ~11.5 and 18.5 Joules. Figure 3a shows that the highest LogEs values tend to decrease with increasing solar activity, which agrees with a suggestion that GCRs, which decrease with solar activity increasing, input to seismic activity. The results in Figures 2, 3 are in agreement with the findings of other authors. For example, in [51] out of phase relationship was found between secular variations in solar activity and seismic energy released on the planet in 1690-2002. The same result was obtained in papers [52, 53]. Also, it was shown in [54] that out of 12 strongest earthquakes (with a magnitude of more than 7.5) that occurred in Japan in 1700-2005, nine events (~70%) were confined to periods of low solar activity, when the intensity of galactic cosmic rays is increased.

*3.2 Increasing seismic activity near magnetic field lines' footprint of newly created radiation belt due to geomagnetic storm*

On March 24, 1991, a powerful geomagnetic storm started at 04:30 UT and reached its negative extremum in the primary phase on March 25, 1991, at 00:30 UT with Dst = - 298 nT. At that time, the CRRES satellite was near L~2.6, and its instruments recorded powerful fluxes of electrons with E ~ 15 MeV and protons with E ~ 20-110 MeV [33]. The MIR orbital station also observed the newly created radiation belt for about two years [34].

On September 1, 2012, a moderate geomagnetic storm started at 22:30 UT and reached its negative extremum in the primary phase on September 3, 2012, at 10:30 UT with Dst = - 69 nT. At that time, the Relativistic Electron Proton Telescope (REPT) on NASA's Van Allen probes board recorded a flux of energetic electrons (3.6 MeV, 4.5 MeV, and 5.6 MeV) at geomagnetic lines 3.0 ≤ L ≤ 3.5. A newly formed belt of relativistic electrons existed for about a month [35] and then was destroyed by the next strong geomagnetic storm on October 1, 2012 [30].

On June 21, 2015, a powerful geomagnetic storm started at 18:30 UT and reached its negative extremum in the primary phase on June 23, 2015, at 04:30 UT with Dst = - 204 nT. At that time, the Magnetic Electron Ion Spectrometer (MagEIS) on NASA's Van Allen probes board recorded a new belt of relativistic electrons with energy E = ~1.06 MeV at geomagnetic lines L ~ 1.5 - 1.8, which was persisted for ~ 11 months [36].

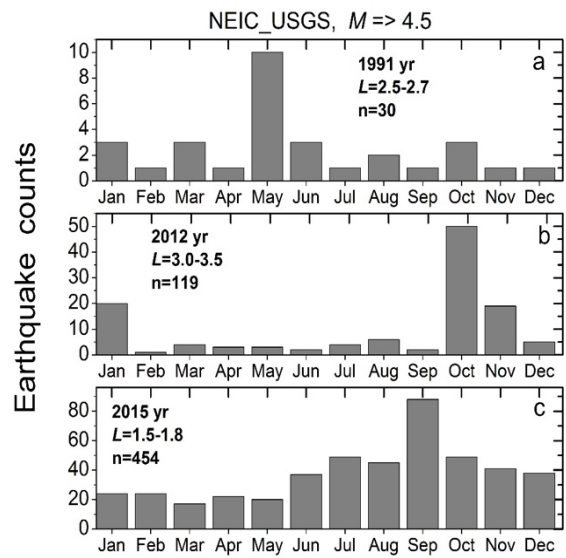

**Figure 4** The histograms of the monthly number of earthquakes with M≥4.5 occurred near the footprints of geomagnetic lines belonging to the newly created belts of relativistic electrons: (a) – 1991, L=2.5-2.7; (b) – 2012, L=3.0-3.5; (c) – 2015, L=1.5-1.8.

Figure 4 presents monthly earthquakes with magnitude M≥4.5 that occurred in 1991, 2012, and 2015 near the footprints of newly created belts of relativistic electrons.

The most significant number of earthquakes occurred in May 1991. The strongest was the M7.0 earthquake in Alaska, which occurred on May 30, 1991, with the coordinates of the epicenter 54.57°N, 161.61°E near the footprint of the geomagnetic line L ~2.69, closely adjacent to the new radiation belt created around L~ 2.6 after a magnetic storm on March 24, 1991 [33,34]. In Figure 4a, the histogram of the number of earthquakes with M≥4.5 near the footprint of the geomagnetic lines L = 2.5 - 2.7 in different months of 1991 is shown. We see that an increase in seismic activity at the base L = 2.5 - 2.7 in May 1991 occurred ~ 2 months after the geomagnetic storm onset.

Figure 4b shows the distribution by months in 2012 of the number of earthquakes with a magnitude of M≥4.5 that occurred near the footprint of the geomagnetic lines L = 3.0 - 3.5, around which a belt of high-energy electrons was formed due to geomagnetic storm on September 1, 2012 [35]. For the base of L = 3.0 - 3.5, the number of earthquakes strongly increased in October 2012; the largest here was an earthquake with M = 7.8, which occurred off the coast of Canada on October 28, 2012, with coordinates 52.79°N, 132.1°W near the footprint of L = ~3.3. This earthquake again happened ~ 2 months after the geomagnetic storm onset on September 1, 2012, which created a storage ring of relativistic electrons.

Figure 4c presents a monthly number of earthquakes with M≥4.5 in 2015, which occurred near the footprint of geomagnetic lines L = 1.5–1.8. A belt of high-energy electrons was formed due to a geomagnetic storm on June 21, 2015 [36]. The increase of seismic activity here started in June 2015, just after a strong geomagnetic storm, but peaked in September 2015. The largest was an earthquake with M = 6.3, which occurred on September 7, 2015, near New Zealand with coordinates of epicenter 32.82°S, 177.86°W in the footprint of L~1.58. The M6.3 events again lag by over two months relative to the magnetic storm onset.

Figure 5 visualizes the correlation between geomagnetic storm onsets and the strongest earthquake in the footprint of magnetic lines belonging to newly created radiation belts.

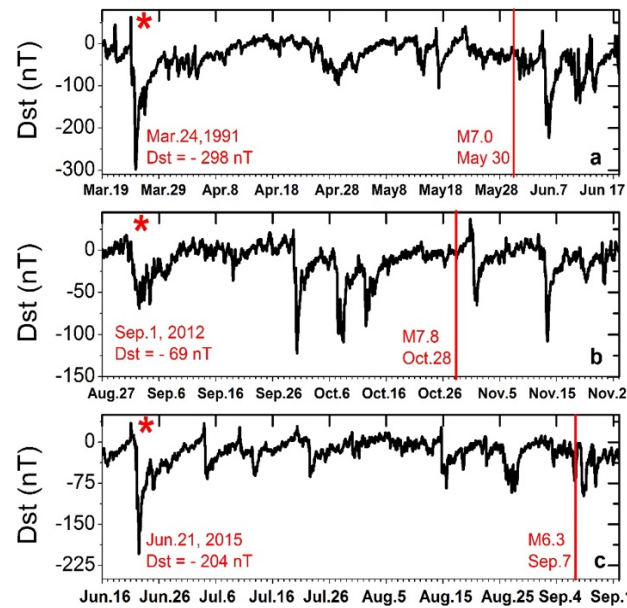

**Figure 5.** Variations of hourly mean Dst-index for three 92-day periods: (a) – 1991, from 19 March, 04:30 UT, (b) - 2012, from 27 August, 22:30 UT, and (c) – 2015, from 16 June, 18:30 UT. Upper red stars mark dates of geomagnetic storm onsets in March 1991; September 2012; and June 2015, and red vertical lines indicate dates of the strongest earthquakes that occurred in the footprint of magnetic lines belonging to the radiation belts newly created in the inner magnetosphere due to indicated magnetic storms: M7.0 in Alaska on May 30, 1991, near the footprint of L~3.3; M7.8 near Canada on October 28, 2012, near the footprint of L~ 2.69; and M6.3 in New Zealand on September 7, 2015, near the footprint of L~1.58.

In Figure 5, the geomagnetic Dst-index from the OMNI database is shown for the three 92-day periods: (a) – 1991, from March 19, 04:30 UT, (b) – 2012, from August 27, 22:30 UT, and (c) – 2015, from June 16, 18:30 UT. Upper red stars mark dates of geomagnetic storm onsets in March 1991; September 2012; and June 2015, and red vertical lines indicate dates of the strongest earthquakes that occurred in the footprint of magnetic lines belonging to the radiation belts newly created due to indicated geomagnetic storms: M7.0 in Alaska on May 30, 1991, near the footprint of L~3.3;   M7.8 near Canada on October 28, 2012, near the footprint of L~ 2.69; and M6.3 in New Zealand on September 7, 2015, near the footprint of L~1.58. The Time delay between storm onset and earthquake occurrence consisted of about 60 days in 1991, 58 days in 2012, and 75 days in 2015; its mean value is equal to ~ 64 days.

Figure 5b shows that the geomagnetic storm on September 1, 2012, was only moderate, but the strong one happened on October 1, 2012, Dst = -122 nT. One may suggest that this stronger storm-induced M7.8 earthquake. It is challenging to find a correlation between the magnetic storm and induced earthquake without an initial idea. Our idea was to investigate seismicity in the footprint of magnetic lines belonging to a new radiation belt created in the inner magnetosphere due to magnetic storms, and this allowed us to get similar results for all three considered cases. The long-living radiation belts are rarely created due to magnetic storms. It so happens that due to the moderate Dst = - 69 nT storm, the new radiation belt was created, but due to the next one strong Dst = -122 storm, it was destroyed [30].

To test if the relationship of seismic activity with the magnetic storm is significant in the spatial domain (near the footprints of magnetic lines belonging to a new radiation belt created due to the storm), we present Figure 6.

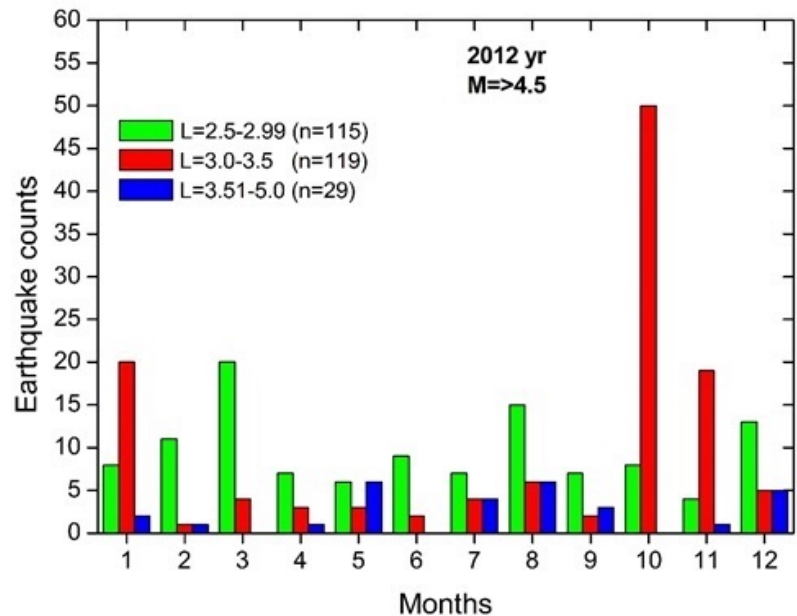

**Figure 6.** Distribution by months of 2012 of the number of earthquakes with magnitude M≥4.5 that occurred at the base of geomagnetic lines L = 2.5 - 2.99 (green), L = 3.0 - 3.5 (red - around which a belt of high-energy electrons was formed after the geomagnetic storm on September 1, 2012), and L = 3.51 - 5.0 (blue).

Figure 6 shows the distribution by months in 2012 of the number of earthquakes with M≥4.5 that occurred near the footprint of L = 3.0 - 3.5 (red bars), as well as in neighboring geomagnetic lines L = 2.5 - 2.99 (green), and L = 3.51 - 5.0 (blue). It is noteworthy that at the base of the lower lines (L = 2.5 - 2.99) and higher (L = 3.51 - 5.0), the distribution of earthquakes by months was more or less uniform. For the base of L = 3.0 - 3.5, around which the belt of high energy particles was formed due to a magnetic storm in September 2012, the number of earthquakes strongly increased ~ 2 months after the onset of the magnetic storm. On the other hand, we presented Figure 6 only as an example to demonstrate the situation for the 2012 year. One cannot expect that the picture will be the same for all possible cases because there are many reasons for earthquake occurrence and, sometimes, they may occur at the base of the nearest magnetic lines, which do not relate to the lines of the newly formed radiation belt. Possibly, this could be the next step of investigations.

A somewhat unexpected result in Figures 4-6 shows that the Time delay between geomagnetic storm onset and earthquake occurrence equals on the average ~64 days. The lag of geomagnetic storms relative to the corpuscular activity of the Sun is about 2-3 days [41], but maybe the lag of seismic activity to the space weather variations can take a relatively long period. For example, the authors of [55] showed that in the territories underlaying by the rocks with low electrical resistivity, the seismic activity increases, on average, 2 - 6 days after geomagnetic storm onset. This suggests that a geomagnetic storm may not trigger an earthquake but rather that the solar wind's energy, a source of magnetic storm generation, can simultaneously be a source of seismic activation.

*3.3 Seismic activity and geometry of the main geomagnetic field*

Some years ago, it was paid attention [56] that spatial scale distribution of earthquake epicenters on the globe is better organized according to geomagnetic coordinates than the geographic ones. It was also revealed that the frequency of earthquake occurrence in different geographical areas depends on the angle of geomagnetic declination (*D*) in these areas. This is evident, for example, from Figure 7, which presents the histogram of several earthquakes with M≥4.5 detected at the globe in 1973-2017 in dependence on *D* value in the epicenter, as estimated with using [40].

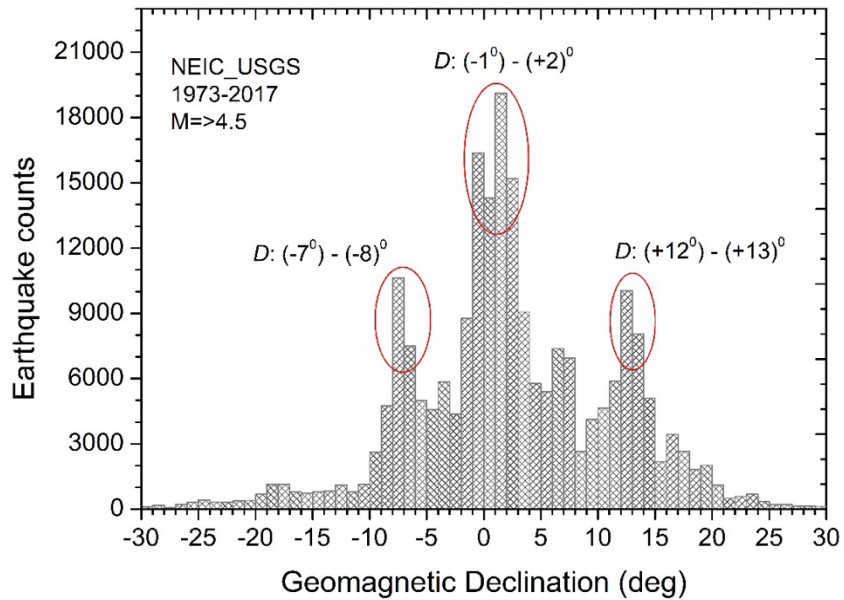

**Figure 7**. Histogram of several earthquakes with M≥4.5 detected at the globe in 1973-2017 in dependence on the angle of geomagnetic declination (*D*) in the epicenter, as estimated with [40].

It is seen from Figure 7 that earthquake occurrence is mainly increased in areas where *D* values are close to zero (central peak). Also, earthquakes occur more often in areas where *D* values are large and positive (right peak) and relatively large and negative (left peak). Figure 8, adapted from [56], shows a spatial location of earthquake epicenters belonging separately to the central peak, right, and left.

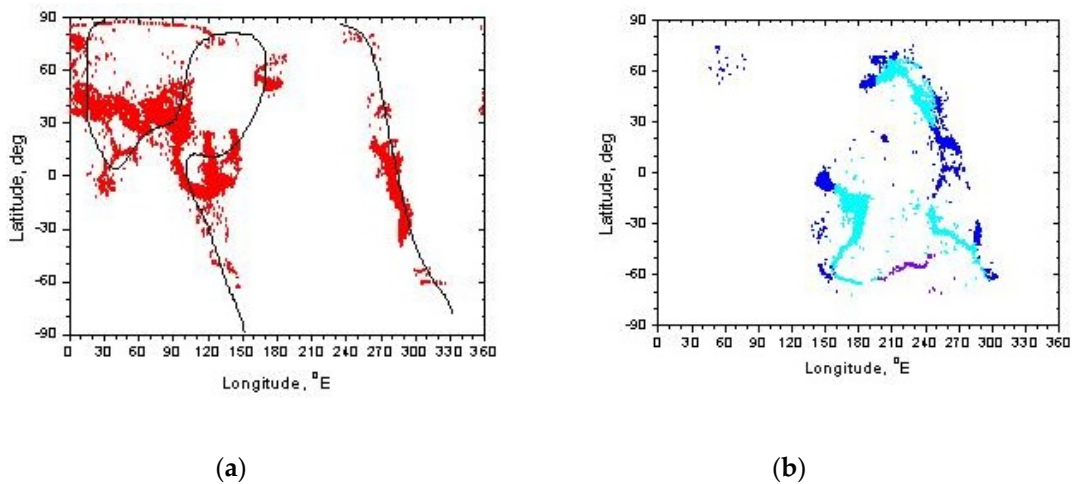

(a)                                      (b)

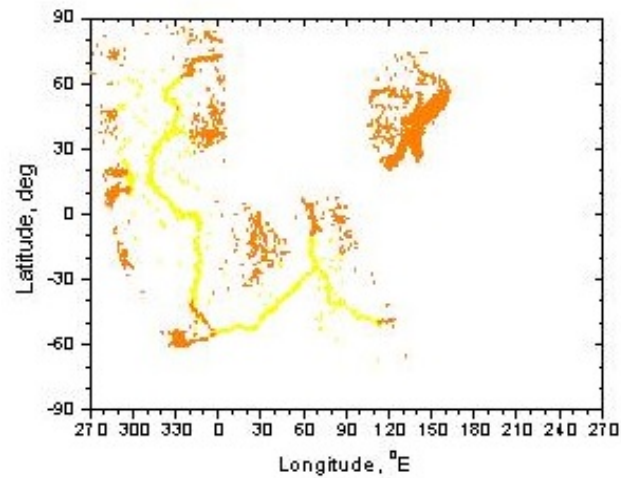

**(c)**

**Figure 8.** The epicenters of earthquakes with M≥4.0 detected at the globe in 1900-2002 and sorted following the angles of geomagnetic declination (*D*) in the epicenters: (a) – epicenters which form the central peak in the function of earthquake counts on *D*, the black lines here show location *D*=0; (b) – epicenters which form a peak at rather large positive *D* values; (c) – epicenters which form a peak at with a rather large negative *D* values, as adopted from [56].

It is not difficult to understand from Figure 8 that epicenters shown in (8a) belong to earthquakes that occurred mainly at the continents in the areas of orogeny, where declination values are close to zero, which is the most evident for the American continent. Epicenters in (8b) belong to earthquakes that occurred mainly at the island arcs (along with the Pacific coast), and epicenters in (8b) belong to earthquakes that occurred mainly in the rift systems at the bottom of the ocean.

It was also noticed that the boundaries of some lithospheric plates are magnetically conjugated, which demonstrates Figures 9, 10 from [57]. In Figure 9, the coordinates of 38 points were determined, distributed relatively evenly along the boundary of the Antarctic lithospheric plate (yellow circles). By using the GEOPACK computational package [40], for each of 38 points, the values of the McIlwain parameter (L) were calculated, and the coordinates of their magnetically conjugate points were determined (red circles).

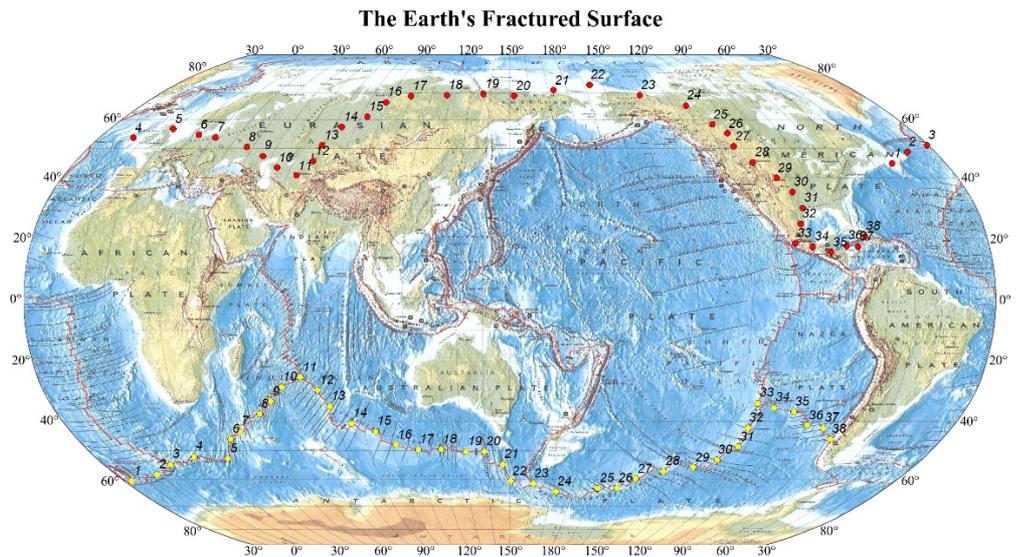

**Figure 9**. Locations of 38 sites in the southern hemisphere along the boundary of the Antarctic lithosphere plate (yellow circles with corresponding numbers) and magnetically conjugated sites in the northern hemisphere (red circles with corresponding numbers) were calculated for geomagnetic conditions of 2000.

Figure 10 shows the L-values of magnetic lines connecting geomagnetically conjugate points (38 pairs) located in the southern and northern hemispheres.

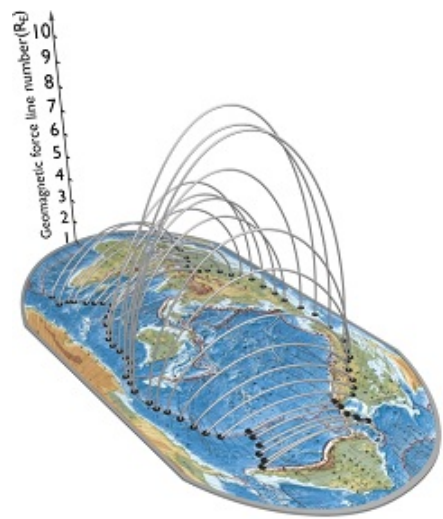

586

**Figure 10.** Geomagnetic field lines connecting magnetically conjugated sites are shown in Figure 9.

Figures 9, 10 show a correspondence between the spatial arrangement of the main tectonic structures and geomagnetic field parameters: the mid-oceanic ridges in the southern hemisphere located along the Antarctic lithospheric plate border which connects with the borders of Pacific, Nazca, South American, Scotia, African, and Australian plate in the southern hemisphere, are being in magnetic conjugation with the zone of the junction of orogenic and platform structures in the northern hemisphere. The effect of geomagnetic conjugation is most clearly manifested between the southern boundary of the Nazca Plate and the northern boundaries of the Cocos and Caribbean plates.

## 4. Discussion and Conclusion

It is well known, the most significant number of earthquakes on the globe occur along the boundaries of the lithospheric plates and in the vicinity of active faults, where the Radon isotopes concentration, which is responsible for ionization and conductivity of the near-ground troposphere, is increased. This allows one to suggest that earthquake occurrence is accompanied by increasing conductivity of air in the lower troposphere.

The released at the globe seismic energy is increased in the low solar activity (Figures 2, 3) when the intensity of the galactic cosmic rays, responsible for ionization and conductivity of the troposphere and lower stratosphere, is increased. These results suggest that earthquake occurrence is accompanied by increasing conductivity of air in the troposphere and lower stratosphere.

Seismic activity increases near the footprint of the geomagnetic field line, belonging to a new (additional) radiation belt created in the lower magnetosphere by the precipitated high energy electrons from the outer radiation belt due to geomagnetic storm (Figures 4-6). As shown in [37], the energetic electrons, which due to geomagnetic storms precipitate from the radiation belt downward, may produce air ionization and thus may increase conductivity in the mesosphere and upper stratosphere. This suggests that earthquake occurrence is accompanied by increasing conductivity of air in the mesosphere and upper stratosphere.

Taking the above into account, one may suggest that increased seismic activity is accompanied by increasing air conductivity in the neutral atmosphere (troposphere-mesosphere), which, in turn, will increase the functioning of the global electric circuit (GEC). It is believed that electromagnetic disturbances in the lithosphere, accompanying the earthquake preparation process, can modify the electric field in the GEC, which results in the appearance of disturbances in the ionosphere [7-11]. In a modified configuration of GEC [18], its external element is located on the magnetopause with the electromotive force generator driven by solar wind energy, and its internal element is located in the solid Earth. This allows one to suggest that a modified GEC [18] may be considered a mediator in transmitting solar wind energy into the Earth's crust.

Distribution of seismic activity at the globe shows a correspondence with the geometry of the Earth's main magnetic field (Figures 7, 8): earthquakes occur more often in areas where the angles of geomagnetic declination are close to zero, that takes place mainly at the continents in the areas of orogeny; also, the peak in the areas with rather large positive D values, that takes place at the island arcs along the Pacific coast, and in the areas with rather a large negative D values, that takes place in the rift systems at the bottom of the ocean. A correspondence is evident between the spatial arrangement of the main tectonic structures and geomagnetic field parameters (Figures 9, 10): the mid-oceanic ridges in the southern hemisphere located along the Antarctic lithospheric plate border are in magnetic conjugation with the zone of the junction of orogenic and platform structures in the northern hemisphere. The effect of geomagnetic conjugation is most clearly manifested between the southern boundary of the Nazca Plate and the northern boundaries of the Cocos and Caribbean plates. On the definition, the magnetic field can influence only electric currents or moving charged particles. Therefore, an observed relation between the distribution of seismic activity and the geometry of the Earth's main magnetic field supports a suggestion that an earthquake has an electrical nature.

The most unexpected and intriguing result in Figures 4-6 is the appearance of an addressed (targeted) strong earthquake in the footprint of a magnetic line belonging to a newly created radiation belt due to a geomagnetic storm. We considered only three long-lived (several weeks and months) radiation belts cases, discussed in [30, 33 -36]. However, geomagnetic storms also generate short-lived radiation belts (some days) [38]. The observed effect of intensification of seismic activity near the footprint of geomagnetic lines of new long-lived radiation belts indicates the advisability of conducting similar studies for short-lived radiation belts, which can be formed after each geomagnetic storm. The approaches developed based on retrospective data for the short-term assessment of increased seismic activity after a geomagnetic storm can then be implemented on the real-time data of the corresponding spacecraft. The experimental base for research can be data from Van Allen probes satellites (2012-2019).

**Supplementary Materials:** There are not supplementary material at this time.

**Author Contributions:** DO and GK provided the concepts of the manuscript. GK organized and wrote the manuscript. All authors provided critical feedback and helped shape the research, analysis, and manuscript. All authors have read and agreed to the published version of the manuscript.

**Funding:** This research received no external funding.

**Institutional Review Board Statement:** Not applicable.

**Informed Consent Statement:** Not applicable.

**Data Availability Statement:** Generated Statement: The original contributions presented in the study are included in the article/supplementary material, further inquiries can be directed to the corresponding author/s.

**Acknowledgments:** Special thanks go to the US Geological Survey and European-Mediterranean Seismological Centre for providing earthquake information services and data. We acknowledge use of NASA/GSFC's Space Physics Data Facility's CDAWeb service, OMNI and Wilcox Solar Observatory data. Special thanks go to the CRRES, and Van Allen Probes scientific missions for providing observations of the radiation belt dynamics, and, especially, to J. B. Blake et al., 1992; D. N. Baker et al., 2004, 2007; S. G. Claudepierre et al., 2017; and R. M. Thorne et al., 2013, who results we used in our investigations. Also, many thanks to the Reviewers.

**Conflicts of Interest:** The authors declare that the research was conducted in the absence of any commercial or financial relationships that could be construed as a potential conflict of interest.